\documentclass[a4paper,12pt]{article}

\usepackage[left=2.5cm,right=2.5cm,top=2.5cm,bottom=2.5cm]{geometry}
\usepackage{graphicx}
\usepackage{amssymb}
\usepackage{multirow}
\usepackage{amsmath}
\usepackage{psfrag}
\usepackage{setspace}
\usepackage{subcaption}
\usepackage[colorlinks=true,bookmarks=true]{hyperref}
\usepackage{float}
\usepackage{cite}
\usepackage[utf8]{inputenc}

\hypersetup{citecolor=blue}
\newcommand{\dif}{\mathrm{d}}
\newcommand{\Eqref}[1]{(\ref{#1})}
\newcommand{\half}{\frac{1}{2}}

\newcommand{\brac}[1]{\left(#1 \right)}
\newcommand{\sbrac}[1]{\left[#1\right]}
\newcommand{\im}{\mathrm{i}}

\newcommand{\Fcal}{\mathcal{F}}
\newcommand{\Acal}{\mathcal{A}}
\newcommand{\Ccal}{\mathcal{C}}

\begin{document}

\title{Charged C-metric in conformal gravity}
\author{Yen-Kheng Lim\footnote{E-mail: phylyk@nus.edu.sg}\\\textit{\normalsize{Department of Physics, National University of Singapore,}}\\\textit{\normalsize{Singapore 117551, Singapore}}}
\date{\normalsize{\today}}
\maketitle

\begin{abstract}
  Using a C-metric-type ansatz, we obtain an exact solution to conformal gravity coupled to a Maxwell electromagnetic field. The solution resembles a C-metric spacetime carrying an electromagnetic charge. The metric is cast in a factorised form which allows us to study the domain structure of its static coordinate regions. This metric reduces to the well-known Mannheim-Kazanas metric under an appropriate limiting procedure, and also reduces to the (Anti-)de Sitter C-metric of Einstein gravity for a particular choice of parameters. 
\end{abstract}


\section{Introduction} \label{intro}

The C-metric is one of the earliest known exact solutions to Einstein gravity, and still many of its features remain relevant for various reasons today. Its compact and elegant form appears almost oblivious to reality as the years unfold as new features and applications have been found with each passing decade. 
\par 
The C-metric was one of the building blocks used to construct the five-dimensional black ring \cite{Emparan:2001wk}, and to provide a description of localised braneworld black holes \cite{Emparan:1999wa,Emparan:1999fd}. In the context of the AdS/CFT correspondence, the C-metric with a negative cosmological constant was used to describe black funnels and droplets \cite{Hubeny:2009ru,Hubeny:2009kz}. Further analysis of its physical properties and causal structure continue to reveal many interesting physics. (See, e.g., Refs.~\cite{Kinnersley:1970zw,Farhoosh:1981kc,Bonnor:1983,Dias:2002mi,Krtous:2003tc,Krtous:2005ej} and related references therein.) Not long ago, Hong and Teo \cite{Hong:2003gx} cast the C-metric in a convenient factorised form in which the solution is parametrised in terms of the roots of its structure functions. Recently, in \cite{Chen:2015vma} this idea has been extended to C-metrics with non-zero cosmological constant.
\par 
With the importance of the C-metric in Einstein gravity, it is natural to study analogous solutions in non-Einstein theories of gravity such as Weyl conformal gravity \cite{Weyl1,Weyl2,Bach1921}. Instead of the usual Einstein-Hilbert action, this formulation of gravity is based on local conformal invariance where the action involves the square of the Weyl tensor. Varying this action results in  fourth-order equations of motion for the metric functions, though in the vacuum case, solutions of Einstein gravity are also vacuum solutions of conformal gravity. 
\par 
Among the most widely used solution in conformal gravity is the spherically symmetric solution obtained by Mannheim and Kazanas (MK) \cite{Mannheim:1988dj}. This solution resembles the Schwarzschild-(A)dS solution, with an additional linear term in its lapse function. The Newtonian limit of these solutions were investigated in \cite{Mannheim:1992tr,Barabash:1999bj}. The charged generalisation of the MK metric was given by Riegert \cite{Riegert:1984zz} and also by Mannheim and Kazanas \cite{1991PhRvD..44..417M} where the rotating generalisation was also given. Other types of solutions were obtained more recently, such as spacetimes with cylindrical symmetry \cite{Brihaye:2009xc,Verbin:2010tq,Said:2012pm}, the Kerr-NUT-(A)dS solution \cite{Liu:2012xn}, and topological black holes \cite{Klemm:1998kf,Cognola:2011nj,Lu:2012xu}. Several solutions have also been studied in theories beyond four-dimensional conformal gravity, such six-dimensional conformal gravity \cite{Lu:2013hx}, and a gravitational action that includes both the Einstein and conformal gravity terms \cite{Lu:2012xu}.
\par 
One of the most promising features of conformal gravity is that it provides a likely explanation of astrophysical phenomena not accounted for in Einsteinian gravity, such as the fitting of galactic rotation curves without the need of introducing dark matter \cite{Mannheim:2010ti,Deliduman:2015vnu}.
Furthermore, the constraints on the parameters obtained from the fitting is also consistent with observations of planetary perihelion precession \cite{Sultana:2012qp}. Further investigations of other experimental tests of gravity are also considered, such as gravitational time delay \cite{Edery:1997hu} and gravitational lensing \cite{Edery:2001at,Sultana:2010zz,Cattani:2013dla,Villanueva:2013gga}.
\par 
The (neutral) C-metric in conformal gravity was studied in detail recently by Meng and Liu in \cite{Meng:2016gyt}. In their paper, the C-metric solution also includes a conformally coupled scalar field. A somewhat similar metric was briefly considered earlier in \cite{Liu:2010sz,Lu:2012xu} where the metric is in the form that is conformal to the Pleba\'{n}ski-Demia\'{n}ski metric.
\par 
In this work, we attempt to derive a solution corresponding to a charged C-metric in conformal gravity and investigate its various properties, in a similar vein to what was previously done for the C-metric in Einstein gravity. In particular, we study the domain structure \cite{Chen:2015vma,Chen:2015zoa} of the solutions which involves analysing the structure of the Lorentzian coordinate regions in a two-dimensional plot.\footnote{The term `\emph{domain structure}' should not be confused with the formalism of the same name in Ref.~\cite{Harmark:2009dh}.} We also aim to show that conformal gravity C-metric contains reduces to the (charged) MK metric under an appropriate limit, similar to how the C-metric reduces to the Reissner-Nordstr\"{o}m solution in Einstein gravity \cite{Mann:1996gj,Hong:2003gx}.
\par 
This paper is organised as follows. In Sec.~\ref{derivation} we present the derivation of the metric using a C-metric-type ansatz and solve the Bach-Maxwell equations describing conformal gravity coupled to an electromagnetic field. Subsequently in Sec.~\ref{sym} we focus on a special choice of parameters that affords various symmetries and provide a convenient form in which one of the structure functions are factorised. In Sec.~\ref{domain} we study the domain structure of the metric and find its possible Lorentzian coordinate regions. Some physical properties of the spacetime are studied in Sec.~\ref{physical}, and various interesting limiting cases of the metric are considered in Sec.~\ref{limits}. This paper ends with some closing remarks in Sec.~\ref{conclusion}.

\section{Derivation of the metric} \label{derivation}
Conformal Weyl gravity is described by the action\footnote{For the expression of the gravitational action we follow the notation of \cite{Riegert:1984zz}, with $(-+++)$ for a Lorentzian signature and a convenient normalisation of the coupling constant to the Maxwell field.}
\begin{align}
 I&=\frac{1}{2\kappa}\int\dif^4x\,\sqrt{-g}\brac{\Ccal_{\mu\nu\rho\sigma}\Ccal^{\mu\nu\rho\sigma}-\Fcal^2},
\end{align}
where $\Ccal$ is the conformal Weyl tensor and $\Fcal=\dif\Acal$ is the Maxwell 2-form flux arising from a 1-form potential $\Acal$. Varying the action with respect to the metric $g$ and $\Acal$ gives the Bach-Maxwell equations
\begin{align}
 W_{\mu\nu}\equiv\brac{2\nabla^\rho\nabla^\sigma+R^{\rho\sigma}}\Ccal_{\mu\rho\sigma\nu}&=2\Fcal_{\mu\lambda}{\Fcal_\nu}^\lambda-\half\Fcal^2 g_{\mu\nu},\label{eom_Bach}\\
                         \nabla_\mu\Fcal^{\mu\nu}&=0. \label{eom_Max}
\end{align}
\par 
We first solve Eq.~\Eqref{eom_Bach} in the vacuum case ($W_{\mu\nu}=0$), beginning with the ansatz
\begin{align}
 \dif s^2&=\frac{1}{(x-y)^2}\brac{Q(y)\dif t^2-\frac{\dif y^2}{Q(y)}+\frac{\dif x^2}{P(x)}+P(x)\dif\phi^2},
\end{align}
where $P(x)$ and $Q(y)$ are functions of only $x$ and $y$, respectively. In the vacuum case, the linear combination $W_{xx}-W_{yy}=0$ leads to
\begin{align}
  PP''''+QQ''''=0,
\end{align}
where primes denote derivatives with respect to their own arguments. This suggests a separation constant $K$ where $PP''''=K=-QQ''''$. Using this separation constant to eliminate the fourth derivatives in $W_{tt}$ and $W_{\phi\phi}$ leads to a single equation,
\begin{align}
 2Q'Q'''-Q''^2&=6K+2P'P'''-P''^2,
\end{align}
which may also be separated with another separation constant $4C$. Solving the resulting third-order ordinary differential equations gives third-order polynomials for $P$ and $Q$ with the requirement that $K=0$. The result is
\begin{align}
 P(x)&=\frac{\brac{p_2^2+C}}{3p_1}x^3+p_2x^2+p_1x+p_0,\nonumber\\
 Q(y)&=\frac{\brac{q_2^2+C}}{3q_1}y^3+q_2y^2+q_1y+q_0,\nonumber
\end{align}
where $p_0,\ldots, p_2$ and $q_0,\ldots,q_2$ are constant coefficients.
\par 
To generalise this solution to include charges, we assume a Maxwell potential that takes the form $\Acal=ey\,\dif t+gx\,\dif\phi$, where $e$ and $g$ respectively denote the electric and magnetic charge parameter. Solving the equations of motion requires a slight modification of the $Q$ polynomial. The result is a nine-parameter metric
\begin{align}
  \dif s^2&=\frac{1}{(x-y)^2}\brac{Q(y)\dif t^2-\frac{\dif y^2}{Q(y)}+\frac{\dif x^2}{P(x)}+P(x)\dif\phi^2},\nonumber\\
  P(x)&=\frac{\brac{p_2^2+C}}{3p_1}x^3+p_2x^2+p_1x+p_0,\nonumber\\
  Q(y)&=\frac{\sbrac{q_2^2+C+3\brac{e^2+g^2}}}{3q_1}y^3+q_2y^2+q_1y+q_0,
\end{align}
which, together with the Maxwell potential $\Acal=ey\,\dif t+gx\,\dif\phi$, solves the Bach-Maxwell equations \Eqref{eom_Bach} and \Eqref{eom_Max}.

\section{Additional symmetries} \label{sym}

For certain special choices of $p_i$ and $q_i$, the metric will carry additional symmetries which allow further simplifications. For example, the solution considered in \cite{Meng:2016gyt} corresponds to the choice
\begin{align}
 C&=\frac{q_2^2p_1-p_2^2q_1}{q_1-p_1},\nonumber\\
 p_2&=\half C_2,\quad q_2=-\half\brac{C_1e_2+C_2},\nonumber\\
 p_1&=C_3,\quad q_1=\half C_1e_2^2+C_2e_2+C_3,\nonumber\\
 p_0&=C_4,\quad q_0=-\brac{\frac{1}{6}C_1e_2^2+\half C_2e_2^2+C_3e_2+C_4}.
\end{align}
In this form, there exists a three-parameter solution which brings $P$ and $Q$ to a form where the neutral solution is characterised by three parameters.
\par 
In this paper, we shall focus our attention to the following choice of parameters:
\begin{align}
 p_1=q_1,\quad |p_2|=|q_2|. \label{choice}
\end{align}
Note that the second condition leads to two possible choices, $p_2=\pm q_2$. We can encode the two distinct choices with $\epsilon=\pm 1$, and upon renaming the other constants, the metric reduces to
\begin{align}
   \dif s^2&=\frac{1}{(x-y)^2}\brac{Q(y)\dif t^2-\frac{\dif y^2}{Q(y)}+\frac{\dif x^2}{P(x)}+P(x)\dif\phi^2},\nonumber\\
     P(x)&=c_0+c_1x+c_2x^2+c_3x^3,\nonumber\\
     Q(y)&=\alpha+c_0+c_1y+\epsilon c_2y^2+\brac{c_3-\frac{e^2+g^2}{c_1}}y^3.\label{conC}
\end{align}
In this form, Eq.~\Eqref{conC} has additional similarities to their counterpart in Einstein gravity which we will explore further in the following sections.
\par
To organise our discussion below, we shall denote the case $\epsilon=1$ as Class I and $\epsilon=-1$ as Class II. One notable feature we see in \Eqref{conC} is that the charge term is qubic, not quartic as in Einstein-Maxwell theory. Therefore, the introduction of charges does not introduce an inner horizon to the spacetime. This is similar to the case of the charged MK solution where the inner horizon is also absent. Furthermore we note another departure from Einstein-Maxwell theory in the relation
\begin{align}
 Q(\xi)-P(\xi)=\alpha+(\epsilon-1)c_2\xi^2-\frac{e^2+g^2}{c_1}\xi^3, \label{PQ_diff}
\end{align}
so that in general, the two structure functions are not identical up to a constant shift. 
\par 
It follows from Eq.~\Eqref{PQ_diff} that in the presence of charges and/or $\epsilon=-1$, the metric does not have the continuous coordinate-translation symmetries enjoyed by its Einstein-Maxwell counterpart. This constrains our ability to fix or eliminate the remaining parameters to cast the metric in a convenient form.
\par 
Nevertheless, we can at least completely factorise one of the structure functions. If we consider factorising $P$, the metric can be reparametrised by introducing
\begin{align}
 c_0=-\mu abc,\quad c_1=\mu(ab+ac+bc),\quad c_2=-\mu(a+b+c),\quad c_3=\mu.
\end{align}
With this parametrisation, Eq.~\Eqref{conC} becomes
\begin{align}
 \dif s^2&=\frac{1}{(x-y)^2}\brac{Q(y)\dif t^2-\frac{\dif y^2}{Q(y)}+\frac{\dif x^2}{P(x)}+P(x)\dif\phi^2},\nonumber\\
	  P(x)&=\mu (x-a)(x-b)(x-c),\nonumber\\
	  Q(y)&=\brac{\mu -\frac{e^2+g^2}{\mu (ab+ac+bc)}}y^3-\epsilon \mu (a+b+c)y^2+\mu (ab+ac+bc)y\nonumber\\
	  &\quad-\mu abc+\alpha, \label{fconC2}
\end{align}
and the Maxwell potential remains unchanged,
\begin{align}
 \Acal=ey\,\dif t+gx\,\dif\phi. \label{Max}
\end{align}
This metric \Eqref{fconC2} and potential \Eqref{Max} will be the form used throughout the rest of this paper. In this form, $P$ is assumed to have real roots.
\par 
In this form, the solution is invariant under the following transformations:
\begin{enumerate}
  \item Rescaling symmetry,
    \begin{align}
     x\rightarrow \lambda x,\quad y&\rightarrow\lambda y,\quad t\rightarrow\lambda t,\quad \phi\rightarrow\lambda\phi,\nonumber\\
     \mu\rightarrow\frac{\mu}{\lambda^3},\quad a&\rightarrow\lambda a,\quad b\rightarrow\lambda b,\quad c\rightarrow\lambda c,\nonumber\\
     e&\rightarrow\frac{e}{\lambda^2},\quad g\rightarrow\frac{g}{\lambda^2},\label{rescaling_sym}
    \end{align}
    for a non-zero, positive constant $\lambda$.
  \item Reflection symmetry,
    \begin{align}
     x&\rightarrow -x,\quad y\rightarrow -y,\quad t\rightarrow-t,\quad \phi\rightarrow-\phi,\nonumber\\
     \mu&\rightarrow-\mu,\quad a\rightarrow-a,\quad b\rightarrow-b,\quad c\rightarrow-c.
    \end{align}

  \item Parameter symmetry,
    \begin{align}
      a\leftrightarrow b,\quad a\leftrightarrow c,\quad b\leftrightarrow c.
    \end{align}

  \item Coordinate symmetry,
    \begin{align}
     x\leftrightarrow y,
    \end{align}
    followed by double-Wick rotations on the pairs $(t,\phi)$ and $(e,g)$,
    \begin{align}
     t&\rightarrow\im\phi,\quad\phi\rightarrow\im t,\nonumber\\
      e&\rightarrow \im g,\quad g\rightarrow \im e.
    \end{align}
\end{enumerate}
Clearly, allowing $\lambda<0$ in the rescaling symmetry \Eqref{rescaling_sym} is equivalent to a positive rescaling followed by a reflection. If we invoke coordinate symmetry on Eq.~\Eqref{fconC2}, we arrive at a form where $Q$ is factorised:
\begin{align}
 \dif s^2&=\frac{1}{(x-y)^2}\brac{Q(y)\dif t^2-\frac{\dif y^2}{Q(y)}+\frac{\dif x^2}{P(x)}+P(x)\dif\phi^2},\nonumber\\
	  P(x)&=\brac{\mu +\frac{e^2+g^2}{\mu (ab+ac+cb)}}x^3-\epsilon \mu (a+b+c)x^2+\mu (ab+ac+bc)x\nonumber\\
	      &\quad-\mu abc-\alpha,\nonumber\\
	  Q(y)&=\mu (y-a)(y-b)(y-c). \label{fconC1}
\end{align}
Therefore we have two alternate forms, \Eqref{fconC2} and \Eqref{fconC1} in which either $P$ or $Q$ is completely factorised. In both cases, the Maxwell potential is still given by Eq.~\Eqref{Max}.
\par 
It should be noted that, in general, the two metrics \Eqref{fconC1} and \Eqref{fconC2} describe different spacetimes. Thus, the analysis of the parameter ranges and domain structure performed below for \Eqref{fconC2} do not automatically apply to the form \Eqref{fconC1}. A separate, but similar analysis should be performed in order to determine the properties of the latter spacetime. 
\par 
The parameter symmetry can be used to fix the ordering of the roots as 
\begin{align}
 a\leq b\leq c. \label{abc_ordering}
\end{align}
We shall also use the reflection symmetry to fix
\begin{align}
 \mu \geq 0. \label{mrange}
\end{align}
With the rescaling symmetry we can fix one of the roots to a particular value. Throughout this paper we will find it convenient to set 
\begin{align}
 c=b+\frac{1}{\mu}, \label{csub}
\end{align}
Note that this choice is consistent with \Eqref{abc_ordering} and \Eqref{mrange}.
\par 
We now have a solution specified by $(\mu,a,b,\alpha,e,g)$, which are four spacetime parameters plus two electromagnetic charges. Altogether, we treat Eq.~\Eqref{fconC2} as a six-parameter solution. 


\section{Coordinate ranges and domain structure} \label{domain}

\subsection{Construction of domain structures in conformal gravity}
Since our metric is described by four spacetime parameters plus two charges, it is not possible to characterise its solutions in a systematic manner using the methods of \cite{Chen:2015vma,Chen:2015zoa}, where the parameter space for (A)dS C-metric is two-dimensional. Furthermore, the fact that the coefficients of $P$ and $Q$ are different leads to many different possible orderings of the roots of $P$ and $Q$.\footnote{This is in stark contrast in the Einstein gravity case, where since $P$ and $Q$ only differ by a constant shift, there are only two possible orderings of the roots. (See, for example, Fig.~1 of Ref.~\cite{Chen:2015zoa}.)}
\par 
Nevertheless we can still consider the possible existence of certain domains by seeking direct numerical examples. We shall briefly review and outline our procedure in this subsection and present the possible domains in Secs.~\ref{ClassI} and \ref{ClassII}. Our method to find the domain structure is as follows.
\par 
The roots of $P$ are already defined in terms of $a$, $b$ and $c=b+1/m$, where we use the symmetries to set $a\leq b\leq c$. Let us denote the roots of $Q$ in increasing order as
\begin{align}
 y_1\leq y_2\leq y_3.
\end{align}
Furthermore, since in \Eqref{fconC2} our electric and magnetic charges only appear in the combination $e^2+g^2$, it will be useful to express the charges as a single quantity
\begin{align}
 q=\sqrt{e^2+g^2},
\end{align}
where we will simply refer to $q$ as the total charge.
\par 
To determine the domain structure for a given set of parameters, one has to first establish the order of these six roots $\{a,b,c,y_1,y_2,y_3 \}$ relative to each other. Knowing the locations of the roots, we would then be able to determine the coordinate ranges where $Q(y)<0$ and $P(x)>0$ which is required for the metric \Eqref{fconC2} to have a Lorentzian $(-+++)$ signature. Plotting these ranges on a two-dimensional plot then gives us the domain structure of the spacetime.
\par 
To demonstrate using a concrete example, let us take $\epsilon=1$, $\mu=1$, $\alpha=0.2$, $q=0.5$, $a=-1$, $b=-0.2$. With these parameters we can easily sketch the curves of $P$ and $Q$ on a common axis using, say, MAPLE or MATHEMATICA.
\begin{figure}
 \begin{center}
  \includegraphics[scale=0.7]{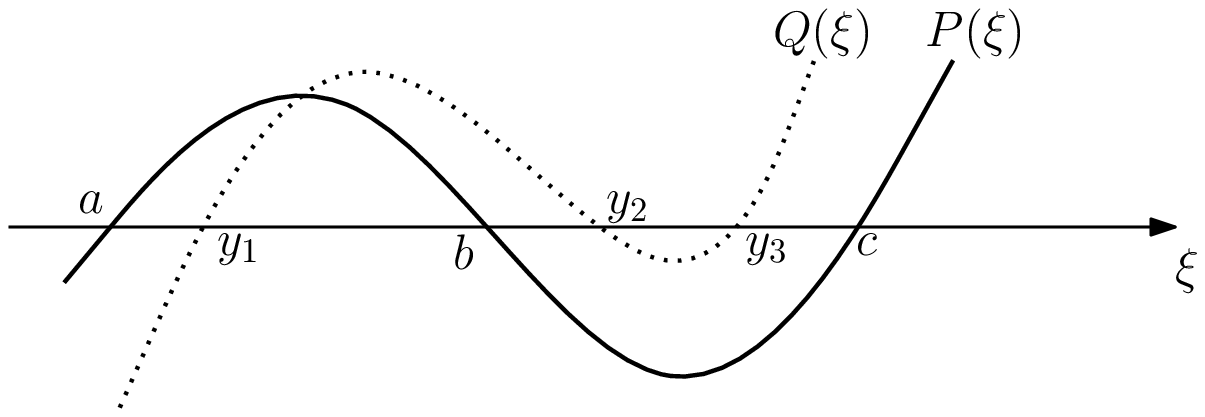}
  \includegraphics[scale=0.8]{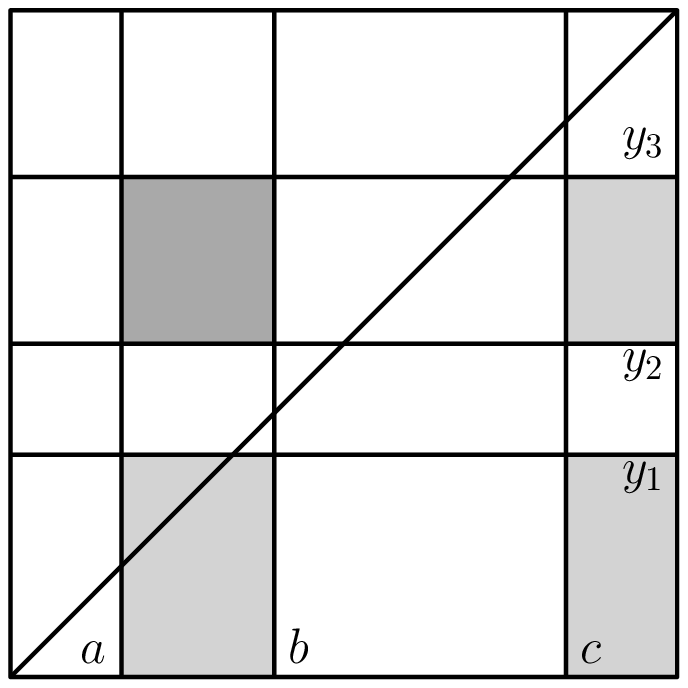}
 \end{center}
 \caption{An example showing the construction of a domain structure for $\epsilon=-1$, $\mu=1$, $\alpha=0.3$, $q=0$, $a=-1$, $b=0.2$, and $c=1$. On the left is the sketch (not to scale) of the functions $P$ (solid) and $Q$ (dotted) showing the ordering of the roots. On the right is the two-dimensional plot where the horizontal (respectively vertical) direction represents the $x$- ($y-$) coordinate. The shaded regions represent the static Lorentzian regions of interest.}
 \label{fig_PQsketch}
\end{figure}
\par 
From the sketch in the left-hand plot of Fig.~\ref{fig_PQsketch}, we can read off the ordering of the roots as
\begin{align}
 a<y_1<b<y_2<y_3<c.
\end{align}
As mentioned above, to have the correct Lorentzian $(-+++)$ signature, we require $P(x)>0$ and $Q(y)<0$. The former is satisfied for the ranges $a<x<b$ and $x>c$, while the latter is satisfied for ranges $y<y_1$ and $y_1<y<y_3$. We then plot the coordinate ranges together on a two-dimensional diagram to find the ranges that satisfy all the required conditions simultaneously. These are shown in the shaded regions in the right-hand plot of Fig.~\ref{fig_PQsketch}. The possible shapes of the shaded regions are what we refer to as the `domain structure'.
\par 
These two-dimensional figures are plots where the horizontal direction represents the $x$-coordinate and the vertical direction represents the $y$-coordinate. The vertical lines represent the symmetry axes ($P=0$) and the horizontal lines represent the horizons ($Q=0$), while the diagonal line is the conformal infinity where $x=y$. The left and right sides of the plots represent $x\rightarrow\pm\infty$, while the upper and lower sides represent $y\rightarrow\pm\infty$. As we will show explicitly in Sec.~\ref{physical}, these limits generally contain curvature singularities. The shaded areas are the static regions of Lorentzian signature, where the darker shade represents areas of particular interest. We are mainly interested in static Lorentzian regions between $a<x<b$, where we will eventually extract the Mannheim-Kazanas spacetime in Sec.~\ref{MKlim} below. Furthermore, our darker-shaded static Lorentzian regions should not include the sides where $x,\,y\rightarrow\pm\infty$ which would correspond to having an observer seeing a naked curvature singularity. 
\par 
Indeed, an observer might pass through horizons to access non-static regions that possibly have curvature singularities. Nevertheless, we wish to view the spacetime from a perspective that is exterior to the black hole. This is partly motivated by physical reasons since, in the MK limit of the metric which will be performed below, the darker-shaded regions are the ones with the most observational significance (for instance, gravitational lensing and other observations mentioned in Sec.~\ref{intro}).

\subsection{Class I: \texorpdfstring{$\epsilon=1$}{epsilon=1}} \label{ClassI}
First we note that, for Class I the two structure functions are related by
\begin{align}
 Q(\xi)-P(\xi)=\alpha-\frac{(e^2+g^2)\xi^3}{\mu(ab+ac+bc)}. \label{PQ_diff_classI}
\end{align}
In the uncharged case the structure functions differ by only a constant shift. We will show in Sec.~\ref{AdSlim} below that the uncharged Class I case is precisely the Einsteinian (A)dS C-metric and has been studied in detail in \cite{Emparan:1999wa,Emparan:1999fd,Hubeny:2009ru,Hubeny:2009kz,Chen:2015vma,Chen:2015zoa}. Therefore we consider cases of non-zero charge unique to conformal gravity. 
\par
\begin{figure}
\begin{center}
 \begin{subfigure}[b]{0.4\textwidth}
  \centering
  \includegraphics[scale=0.8]{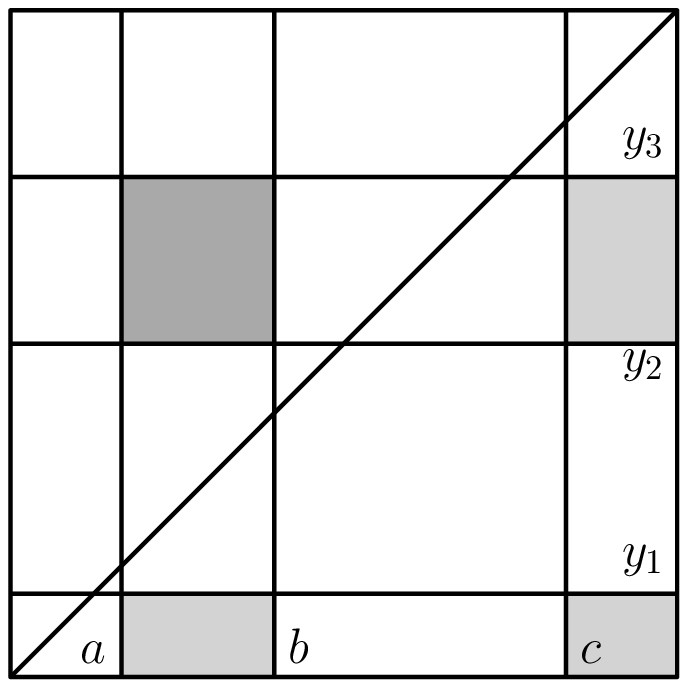}
  \caption{$\alpha=0.3$, $q=0.1$.}
  \label{fig_rangeI-01}
 \end{subfigure}
 \hspace{0.2cm}
 \begin{subfigure}[b]{0.4\textwidth}
  \centering
  \includegraphics[scale=0.8]{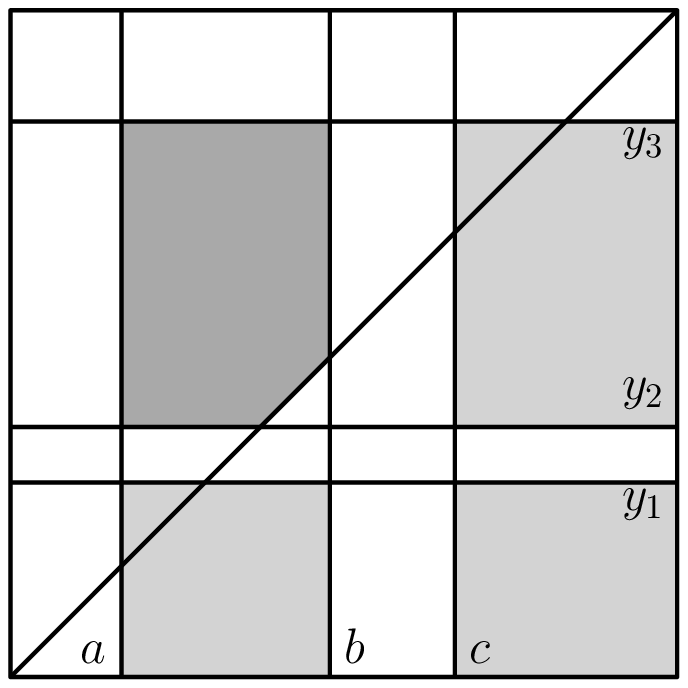}
  \caption{$\alpha=-0.2$, $q=0.1$.}
  \label{fig_rangeI-02}
 \end{subfigure}
 \\
 \begin{subfigure}[b]{0.4\textwidth}
  \centering
  \includegraphics[scale=0.8]{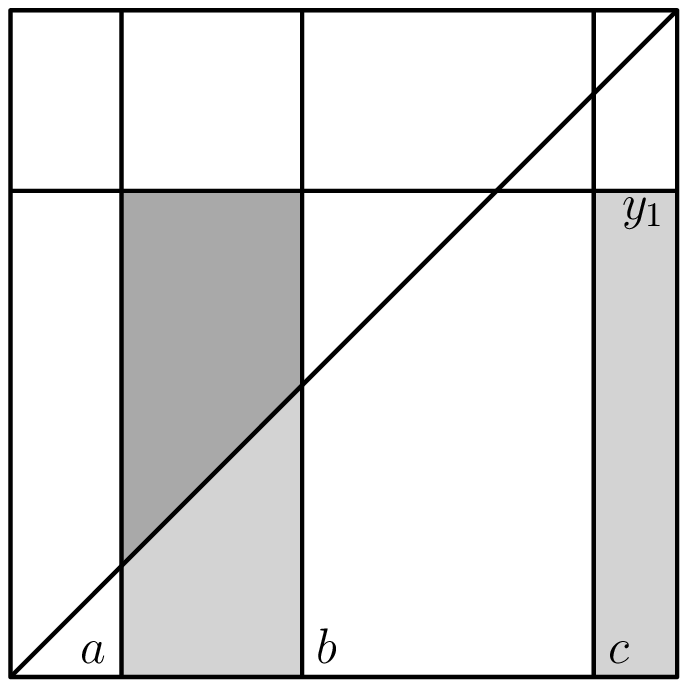}
  \caption{$\alpha=-0.2$, $q=0.6$.}
  \label{fig_rangeI-03}
 \end{subfigure}
 \hspace{0.2cm}
 \begin{subfigure}[b]{0.4\textwidth}
  \centering
  \includegraphics[scale=0.8]{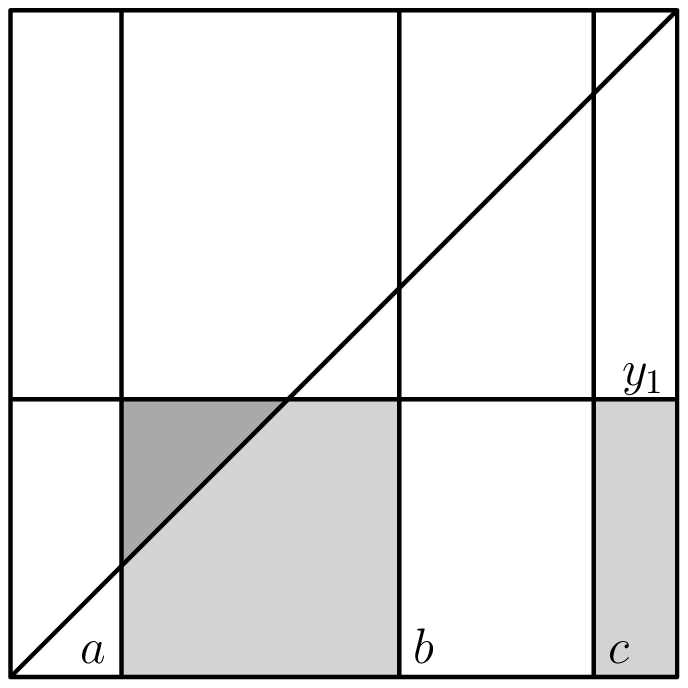}
  \caption{$\alpha=0.5$, $q=0.8$.}
  \label{fig_rangeI-04}
 \end{subfigure}
 \\
 \begin{subfigure}[b]{0.4\textwidth}
  \centering
  \includegraphics[scale=0.8]{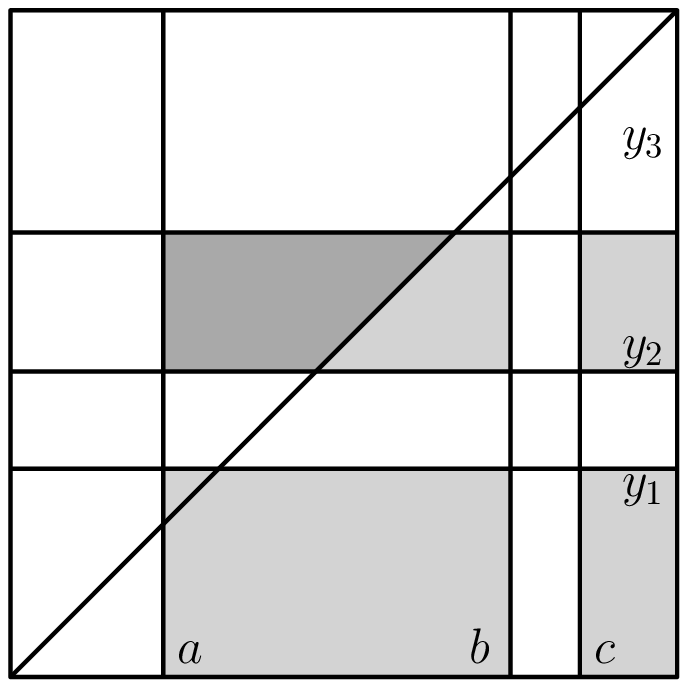}
  \caption{$\alpha=-0.95$, $q=1$.}
  \label{fig_rangeI-05}
 \end{subfigure}
\end{center}
\caption{Possible Class I domain structures for $\mu=1$ and various values of $\alpha$ and $q$. For Figs.~\ref{fig_rangeI-01}-\ref{fig_rangeI-04} the roots are chosen to be $a=-1$, $b=-0.2$ and $c=1$, while for Fig.~\ref{fig_rangeI-05} the roots are $a=-1$ and $b=1$. The shaded regions correspond to static regions with Lorentzian signature $(-+++)$, one of which is the region of our primary interest that is shaded in dark gray. The diagonal line represents the conformal infinity $x=y$.}
\label{fig_rangeI}
\end{figure}

 
By checking various numerical values of the metric parameters, we obtain the possible domain structures shown in Fig.~\ref{fig_rangeI}. We find the same five possible shapes that were present in the (A)dS C-metric in Einstein gravity, namely the square box, `chipped' box (a box with a corner cut off by conformal infinity), vertical trapezium, triangle, and horizontal trapezium. 

\par 
Figure \ref{fig_rangeI-01} shows a square box which is analogous to the de Sitter C-metric considered in \cite{Chen:2015vma}. It corresponds to a Lorentzian region bounded by two symmetry axes $x=a$ and $x=b$, and two horizons $y=y_2$ and $y=y_3$. From the perspective of an observer in this square box, the horizon $y=y_3$ conceals the curvature singularity at $y\rightarrow\infty$. Therefore we shall interpret $y=y_3$ as the black hole horizon. This horizon has a finite area, extending from one symmetry axes at $x=a$ to the other at $x=a$. Loosely speaking, we may say that this black hole horizon has a spherical topology. The second horizon is located at $y=y_2$, which is also finite and it conceals the observer from the conformal infinity, thus we shall refer to it as an acceleration, or cosmological horizon.
\par
The `chipped box' and vertical trapezium in Figs.~\ref{fig_rangeI-02} and \ref{fig_rangeI-03} respectively shows similarly finite black-hole horizons of spherical topology. For the `chipped' box, the second horizon at $y=y_2$ intersects the diagonal line $x=y$. Therefore the acceleration/cosmological horizon extends all the way `to conformal infinity', and does not intersect the second symmetry axis. Such boxes in Einstein gravity were interpreted as the `fast' accelerating AdS C-metrics, where the acceleration parameter exceeds the AdS curvature parameter, i.e., $A>\frac{1}{\ell}$ \cite{Dias:2002mi,Chen:2015vma}. For the vertical trapeziums there is no second horizon in the Lorentzian region; this is the analogue of the `slow' acceleration case $A<\frac{1}{\ell}$ in Einstein gravity.
\par 
The triangle and vertical trapezium of Figs.~\ref{fig_rangeI-04} and \ref{fig_rangeI-05} contain black hole horizons that extend to conformal infinity and intersect only one symmetry axis. Thus we conclude that the horizon is infinite in extent and has the domain structure similar to the deformed hyperbolic black holes in Einstein gravity \cite{Chen:2015zoa}.

\subsection{Class II: \texorpdfstring{$\epsilon=-1$}{epsilon=-1}} \label{ClassII}
Proceeding to Class II solutions, for $\epsilon=-1$ the structure functions are related by
\begin{align}
 Q(\xi)-P(\xi)=\alpha+2\mu(a+b+c)\xi^2-\frac{(e^2+g^2)\xi^3}{\mu(ab+ac+bc)}. \label{PQ_diff_classII}
\end{align}
Thus we see that the situation in Class II is more complicated, difference between $P$ and $Q$ also contains a quadratic term. It follows that there are three possible intersection points between the two structure functions. In uncharged case, the difference in \Eqref{PQ_diff_classII} is only quadratic, and only leads to two distinct intersection points when $\alpha$ is non-zero.
\begin{figure}
\begin{center}
 \begin{subfigure}[b]{0.4\textwidth}
  \centering
  \includegraphics[scale=0.8]{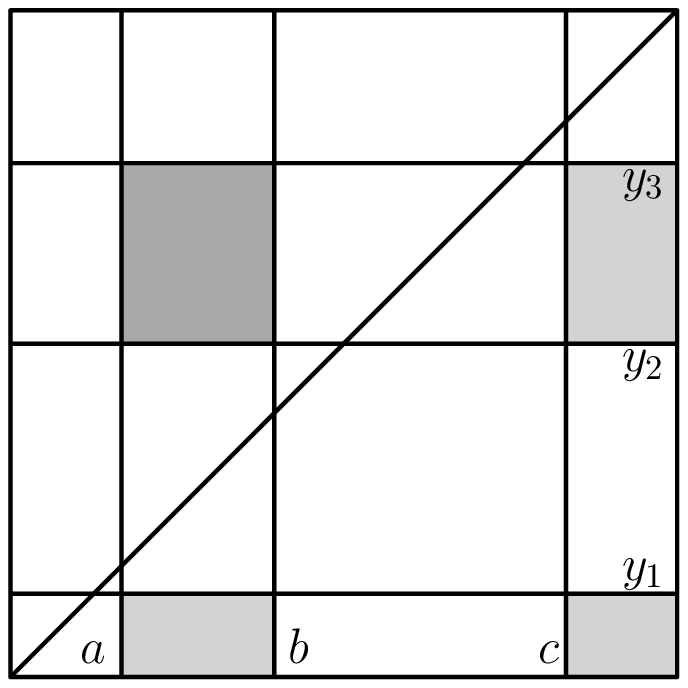}
  \caption{$\alpha=0.05$, $q=0$.}
  \label{fig_rangeII-01}
 \end{subfigure}
 \hspace{0.2cm}
 \begin{subfigure}[b]{0.4\textwidth}
  \centering
  \includegraphics[scale=0.8]{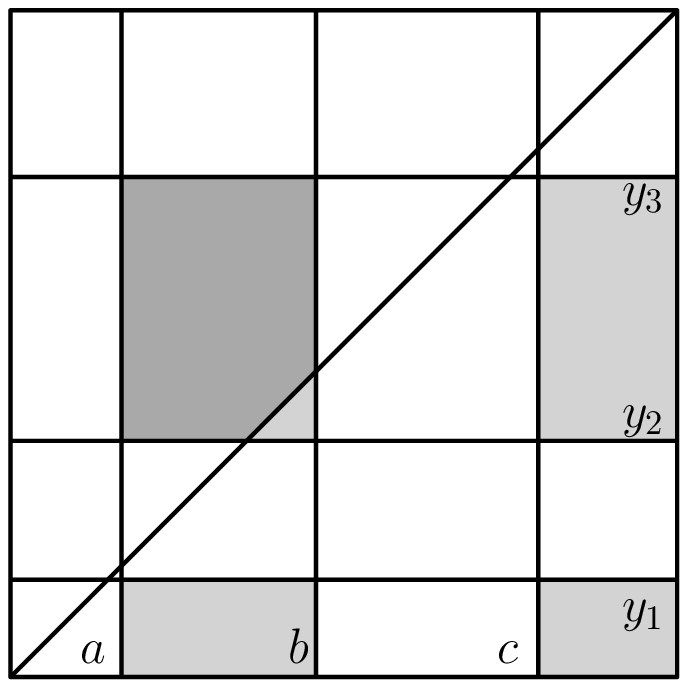}
  \caption{$\alpha=-0.6$, $q=0$.}
  \label{fig_rangeII-02}
 \end{subfigure}
 \\
 \begin{subfigure}[b]{0.4\textwidth}
  \centering
  \includegraphics[scale=0.8]{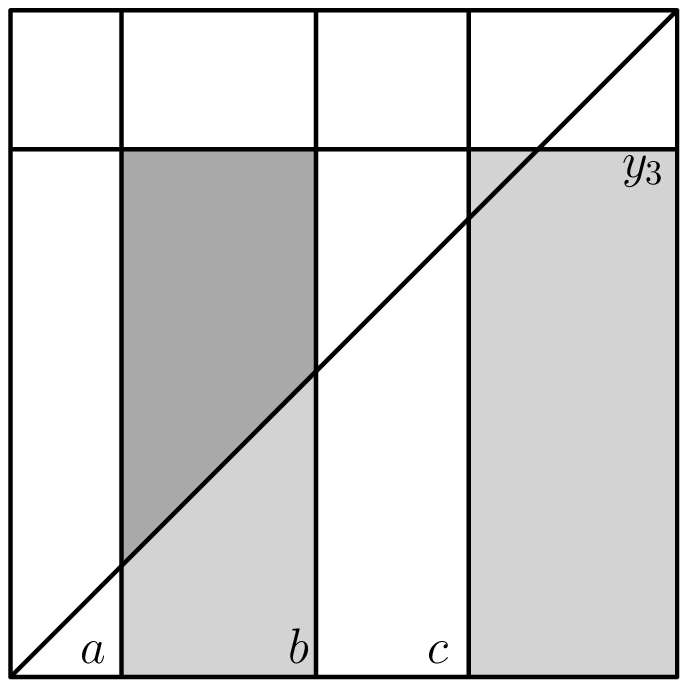}
  \caption{$\alpha=-1.5$, $q=0$.}
  \label{fig_rangeII-03}
 \end{subfigure}
 \hspace{0.2cm}
 \begin{subfigure}[b]{0.4\textwidth}
  \centering
  \includegraphics[scale=0.8]{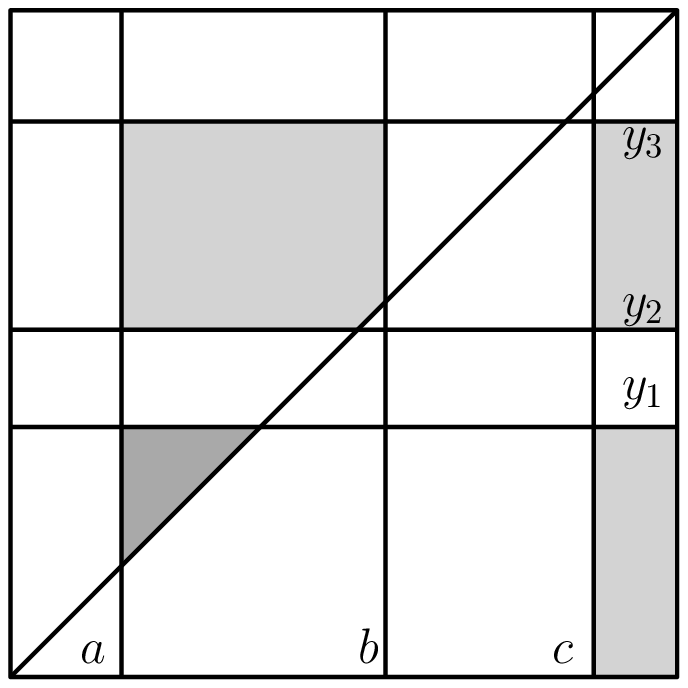}
  \caption{$\alpha=-0.7$, $q=0.5$.}
  \label{fig_rangeII-04}
 \end{subfigure}
 \\
 \begin{subfigure}[b]{0.4\textwidth}
  \centering
  \includegraphics[scale=0.8]{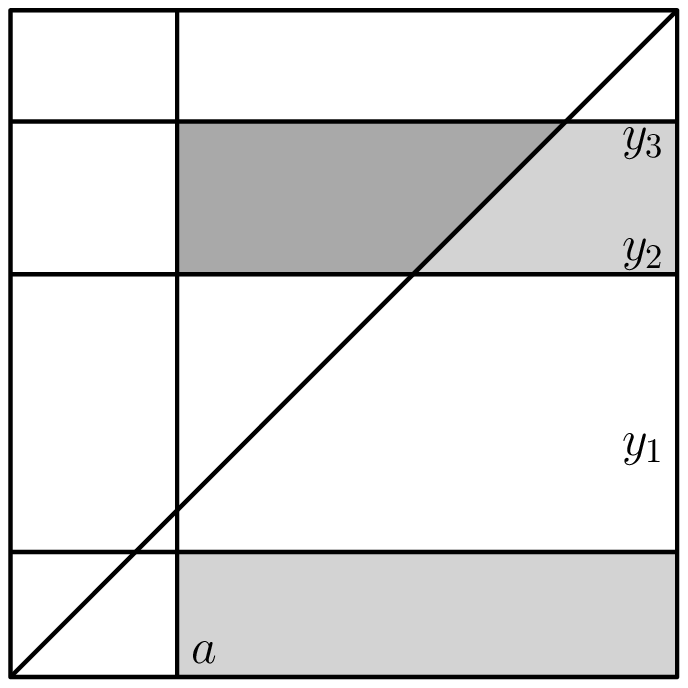}
  \caption{$\alpha=-3$, $q=0$.}
  \label{fig_rangeII-05}
 \end{subfigure}
\end{center}
\caption{Possible Class II domain structures for $\mu=1$ and various values of $\alpha$ and $q$. For Figs.~\ref{fig_rangeII-01}-\ref{fig_rangeII-04} the roots are chosen to be $a=-1$ and $b=0.2$, while for Fig.~\ref{fig_rangeII-05} the roots are $a=-1$ and $b=1$. The shaded regions correspond to static regions with Lorentzian signature $(-+++)$, one of which is the region of our primary interest that is shaded in dark gray. The diagonal line represents the conformal infinity $x=y$.}
\label{fig_rangeII}
\end{figure}
\par 
Seeking out various numerical examples, we find the possible domain structures in Fig.~\ref{fig_rangeII}. We find the same possible domains as in Class I. Thus we conclude that in general, Class I and Class II are physically similar in terms of the horizon configurations and the symmetry axes. The distinction between Class I and II, as we will discuss in Sec.~\ref{limits}, lies in the uncharged case $e=g=0$. In the uncharged case Class I immediately reduces to the Einsteinian C-metric while Class II does not, except for a specific choice of parameters.

\section{Physical properties} \label{physical}
Our domains of interest lie between $a\leq x\leq b$ where the boundaries are the symmetry axes where $P=0$. For a given periodicity of the angular coordinate $\phi$,  the conical deficit at these axes can be calculated as
\begin{align}
 \delta_{i}=2\pi-\kappa_{\mathrm{E}i}\Delta\phi,
\end{align}
where $i=a,b$ and $\kappa_{\mathrm{E}}$ is the Euclidean surface gravity \cite{Chen:2010zu}, or, the ratio between the circumference and the radius of an infinitesimally small circle around the respective axes. For our metric \Eqref{fconC2}, they are given by
\begin{align}
 \kappa_{\mathrm{E}a}&=\half\left|P'(a) \right|=\half \mu(b-a)(c-a), \label{kappa_Ea}\\
 \kappa_{\mathrm{E}b}&=\half\left|P'(b) \right|=\half \mu(b-a)(b-c). \label{kappa_Eb}
\end{align}
We can remove one of the two conical singularities by appropriately fixing the periodicity $\Delta\phi$. The two possible choices are
\begin{align}
 \Delta\phi&=\frac{2\pi}{\kappa_{\mathrm{E}a}}:\quad\delta_a=0,\quad\delta_b=2\pi\frac{b-a}{c-a}, \label{delta_a}\\
 \Delta\phi&=\frac{2\pi}{\kappa_{\mathrm{E}b}}:\quad\delta_a=-2\pi\frac{b-a}{c-b},\quad\delta_b=0. \label{delta_b}
\end{align}
Therefore, we see that the first choice removes the conical singularity at $x=a$, leaving a conical \emph{excess} at $x=b$, ($\delta_b>0$) which we regard as a cosmic strut pushing against the black hole, while the second choice removes the singularity at $x=b$ but leaves a conical \emph{deficit} at $x=a$, ($\delta_a<0$) which is regarded as a cosmic string pulling the black hole. (See, e.g., \cite{Hong:2003gx,Griffiths:2006tk,Griffiths:2009dfa} and references therein.) In either case, we have the interpretation that the black hole is being accelerated along the $x=a$ axis.
\par
Next we consider the curvature invariants of the spacetime. The Kretschmann invariant is, for Class I,
\begin{align}
 R_{\mu\nu\rho\sigma}R^{\mu\nu\rho\sigma}&=24\alpha^2+12\mu ^2(x-y)^6+\frac{24y(e^2+g^2)}{\mu (ab+ca+cb)}\brac{\mu (x-y)^5-\alpha x(x+y)}\nonumber\\
 &\quad+\frac{12\brac{e^4+g^4}}{\mu ^2(ab+ac+cb)^2}\brac{3x^3+y^4-6x^3y-4xy^3+8x^2y^2}, \label{ClassIKret}
\end{align}
so in Class I, if either of $\mu$, $e$ or $g$ are non-zero, there are curvature singularities for $x,\,y,\rightarrow\pm\infty$. Therefore in these cases, the outermost edges of Figs.~\ref{fig_rangeI} and \ref{fig_rangeII} represent a curvature singularity.
\par 
As mentioned in Sec.~\ref{domain}, the uncharged case of Class I reduces to the (A)dS C-metric of Einstein gravity. We also can see this here if we put $e=g=0$ in Eq.~\Eqref{ClassIKret}, the curvature invariant simply becomes $R_{\mu\nu\rho\sigma}R^{\mu\nu\rho\sigma}=24\alpha^2+12\mu ^2(x-y)^6$. Comparing this to the Kretschmann invariant of the (A)dS C-metric in Einstein gravity, we see that $\mu$ plays the role of the `mass' parameter, where its vanishing leaves us with an empty, constant-curvature spacetime.
\par 
The Kretschmann invariant for Class II is more complicated:
\begin{align}
 R_{\mu\nu\rho\sigma}R^{\mu\nu\rho\sigma}&=24\alpha^2+16\alpha \mu (a+b+c)(x^2+y^2+4xy)+\mu ^2\Bigl(12x^6-16x^5b+16x^4b^2\nonumber\\
     &\quad -16ax^5+16a^2x^4-16x^5c+16x^4c^2+12y^6+180x^4y^2-240x^3y^3-72x^5y\nonumber\\
     &\quad +16y^4a^2+16y^4b^2+16y^4c^2+16y^5a+16y^5b+16y^5c-72xy^5+180x^2y^4\nonumber\\
     &\quad +128y^2cax^2+128y^2abx^2+128y^2cbx^2+64y^2x^2c^2+64y^2x^2b^2+64y^2a^2x^2\nonumber\\
     &\quad+160y^3ax^2+32y^4ca+32y^4ab+32y^4cb-160x^3y^2a-160x^3y^2b\nonumber\\
     &\quad-160x^3y^2c+80x^4yb+80x^4ya-80xy^4a-80xy^4b+80x^4yc-80xy^4c\nonumber\\
     &\quad+160y^3cx^2+160y^3bx^2+32ax^4c+32ax^4b+32x^4bc\Bigr)\nonumber\\
     &\quad-\frac{8y(e^2+g^2)}{\mu (ab+ac+bc)}\Bigl[3x\alpha (x+y)+\mu \Bigl(6x^4a+2ay^4-6yax^3+14y^2ax^2\nonumber\\
     &\quad-4y^3ax+6x^4b+2y^4b-6ybx^3+14y^2bx^2-4y^3xb-15xy^4+15x^4y\nonumber\\
     &\quad-6ycx^3-4xy^3c+3y^5+14y^2cx^2-3x^5+30y^3x^2+2y^4c\nonumber\\
     &\quad-30y^2x^3+6x^4c\Bigl)\Bigl]\nonumber\\
     &\quad+\frac{12y^2(e^4+g^4)}{\mu ^2(ab+ac+cb)^2}\brac{3x^4-6x^3y+8x^2y^2-4xy^3+y^4}. \label{ClassIIKret}
\end{align}
Nevertheless, we have a similar result that in general, there exist curvature singularities at $x,\,y\rightarrow\pm\infty$.

\section{Limiting cases} \label{limits}

\subsection{Mannheim-Kazanas metric} \label{MKlim}

For a spacetime with a domain structure bounded by two symmetry axes $x=a$ and $x=b$, we have pointed out in Sec.~\ref{physical} that one cannot simultaneously remove both conical singularities by fixing an appropriate periodicity of $\phi$. Upon removal of a conical singularity at one axis, the other has either a conical excess or deficit given in Eqs.~\Eqref{delta_a} or \Eqref{delta_b}. Nevertheless, we see from these two equations that in both cases, $\delta_a$ and $\delta_b$ can be rendered simultaneously zero if $a\rightarrow b$.
\par 
However, this entails shrinking the coordinate range $a<x<b$ to zero unless we scale $x$ accordingly. To ensure our coordinates are well defined in this limit, we introduce the transformation
\begin{align}
 x=b-\half\delta\brac{\cos\theta+1},\quad y=b+\frac{1}{r},\quad \phi=\frac{2\varphi}{\delta},\quad a=b-\delta.
\end{align}
Substituting this into the Class I ($\epsilon=1$) case of \Eqref{fconC2}, and taking the limit $\delta\rightarrow 0$, we obtain
\begin{align}
 \dif s^2&=-f(r)\dif t^2+f(r)^{-1}\dif r^2+r^2\brac{\dif\theta^2+\sin^2\theta\,\dif\varphi^2},\nonumber\\
      f(r)&=w+\frac{u}{r}+vr-k r^2,\label{nonaccel_sph}
\end{align}
where $u$, $v$, $w$ and $k$ are given by
\begin{align}
 u&=\frac{e^2+g^2-2\mu b-3\mu ^2b^2}{b(3\mu b+2)},\nonumber\\
 v&=\frac{3b(e^2+g^2)}{2+3\mu b},\nonumber\\
 w&=\frac{2+3\mu b+3(e^2+g^2)}{2+3\mu b},\nonumber\\
 k&=\frac{2\alpha+3\alpha \mu b-b^2(e^2+g^2)}{2+3\mu b}. \label{nonaccelI}
\end{align}
The resulting Maxwell potential, up to an irrelevant constant term, is
\begin{align}
 \Acal&=\frac{e}{r}\,\dif t+ g\cos\theta\,\dif\phi, \label{nonaccel_Max}
\end{align}
We can easily check that the parameters defined in \Eqref{nonaccelI} satisfy
\begin{align}
 w^2-1-3uv=3\brac{e^2+g^2}, \label{MK_constraint}
\end{align}
showing that this is the charged black hole in conformal gravity \cite{Riegert:1984zz,1991PhRvD..44..417M}, albeit with different parametrisation.
\par 
In the uncharged case, the reduction to the Schawrzschild-(A)dS can be seen by putting $e=g=0$ in \Eqref{nonaccelI}, the solution reduces to
\begin{align}
 \dif s^2&=-f(r)\dif t^2+f(r)^{-1}\dif r^2+r^2\brac{\dif\theta^2+\sin^2\theta\,\dif\varphi^2},\nonumber\\
      f(r)&=1-\frac{\mu}{r}-\alpha r^2,
\end{align}
corresponding to the Schwarzschild-(A)dS solution with mass parameter $\mu=2m$ and curvature parameter $\alpha=-\frac{1}{\ell^2}=\frac{\Lambda}{3}$.
\par 
If we apply this limiting procedure to Class II with $\epsilon=-1$, we obtain the same form as \Eqref{nonaccel_sph}, but with different coefficients of $r$:
\begin{align}
 u&=\frac{e^2+g^2-2\mu b-3\mu ^2b^2}{b(2+3\mu b)},\nonumber\\
 v&=\frac{b\sbrac{3(e^2+g^2)-8-36\mu ^2b^2-36\mu b}}{2+3\mu b},\nonumber\\
 w&=\frac{3(e^2+g^2)-2-15\mu b-18\mu ^2b^2}{2+3\mu b},\nonumber\\
 k&=\frac{18m^2b^4+18mb^3+2\alpha+3\alpha mb+4b^2-b^2(e^2+g^2)}{2+3\mu b}, \label{nonaccelII}
\end{align}
where they also satisfy Eq.~\Eqref{nonaccelII}. This is again the charged Mannheim-Kazanas spacetime with yet another parametrisation.
\par 
For the uncharged case, taking $e=g=0$ in Eq.~\Eqref{nonaccelII} and further identifying
\begin{align}
 m=\beta\brac{2-3\beta\gamma},\quad b=-\frac{1}{6\beta},\quad\alpha=k-\frac{\gamma}{12\beta},
\end{align}
we see that  \Eqref{nonaccel_sph} reduces to 
\begin{align}
 \dif s^2&=-f(r)\dif t^2+f(r)^{-1}\dif r^2+r^2\brac{\dif\theta^2+\sin^2\theta\,\dif\varphi^2},\nonumber\\
      f(r)&=1-3\beta\gamma-\frac{2(\beta-3\beta\gamma)}{r}+\gamma r-k r^2,
\end{align}
which is precisely the well-known Mannheim-Kazanas vacuum solution \cite{Mannheim:1988dj}.

\subsection{(A)dS C-metric} \label{AdSlim}
As mentioned above, the main feature of the conformal gravity C-metric that distinguishes it from its Einsteinian counterpart can be traced to the fact that $Q(\xi)-P(\xi)$ is not equal to a constant. Nevertheless, by inspection of Eqs.~\Eqref{PQ_diff}, \Eqref{PQ_diff_classI} or \Eqref{PQ_diff_classII} the difference can be equal to constant $\alpha$ by a suitable choice of parameters.
\par 
To remove the cubic term from $Q(\xi)-P(\xi)$, we require $e=g=0$. Then, the entire Class I metric with $\epsilon=1$ satisfies 
\begin{align}
 R_{\mu\nu}=3\alpha g_{\mu\nu}, \label{EinsteinGrav}
\end{align}
showing that it is a solution to Einstein's equation with cosmological constant $\Lambda=3\alpha$.
\par 
For Class II, Eq.~\Eqref{PQ_diff_classII} tells us that $Q(\xi)-P(\xi)$ can be made constant by setting $m(a+b+c)=0$ in addition to $e=g=0$. Recalling \Eqref{csub}, the former condition is equivalent to
\begin{align}
 a+2b+\frac{1}{m}=0.
\end{align}

\section{Conclusion}\label{conclusion}

In this paper we have attempted to derive a charged C-metric-type solution in conformal gravity. Starting with an anstaz that resembles the C-metric in Einstein gravity, we obtained a nine-parameter solution to the Bach-Maxwell equations. By construction, two of these parameters are the electric and magnetic charges, though at this stage, we have no reason to conclude that all the remaining seven parameters carry physical significance, as some of them are possibly kinematical parameters.
\par 
We have focused our attention to a six-parameter subset of the solution. The motivation for doing so is two-fold. First, this subset contains some additional symmetries and it allows us to rewrite one of the structure functions in a convenient factorised form. Secondly, this choice is inspired by the analogy that the charged Einsteinian C-metric is a one-parameter generalisation of the Reissner-Nordstr\"{o}m solution, hence we expect a conformal gravity C-metric to also be a one-parameter generalisation of the charged Mannheim-Kazanas solution. Since only one of the structure functions are fully factorised, we obtain the possible domain structures of the C-metric by directly searching for numerical examples. Our charged C-metrics contain five possible domain shapes that are similar to those in the neutral (A)dS C-metric in Einstein gravity. 
\par 
In this paper we have mostly confined ourselves within one static Lorentzian region of interest. A further exploration of the metric can be done by extending across the horizons into different regions to study its global and causal structure. Since this requires extension of the spacetime across its horizons, it is probably more convenient to use the form given in Eq.~\Eqref{fconC1} instead of \Eqref{fconC2}. Furthermore, we have only considered a specific choice of parameters as given in Eq.~\Eqref{choice}. It would be interesting to explore other parameter choices in further detail, for instance a choice that contains the solution described by \cite{Meng:2016gyt}.
\par 
It would also be interesting to consider null and time-like geodesics for this spacetime. In the spherically symmetric case of the Mannheim-Kazanas metric, it was shown in \cite{Edery:1997hu} that conformal gravity affects time-like and null geodesics very differently from Einstein gravity. Thus it would be interesting to see its corresponding cases for the C-metric. Furthermore, since solutions to the Bach-Maxwell equations are conformally invariant, it might be worth studying a metric with a gauge in which the overall conformal factor $(x-y)^{-2}$ is removed, for example, one of the gauges considered in \cite{Meng:2016gyt}. For metrics of this form the geodesic equations of time-like particles would possibly be separable, as it is this factor that originally prevented the separation of the time-like geodesic equations of the Einsteinian C-metric.

\section*{Acknowledgement}
The author would like to thank Qinghai Wang whose collaboration on a different work inspired the present one.

\bibliographystyle{mybib}

\bibliography{myrefs}

\providecommand{\href}[2]{#2}\begingroup\raggedright\begin{thebibliography}{10}

\bibitem{Emparan:2001wk}
R.~Emparan and H.~S. Reall, {\it {`Generalized Weyl solutions'}},  Phys.~Rev.~D
  {\bf 65} (2002) 084025, [\href{http://arxiv.org/abs/hep-th/0110258}{{\tt
  hep-th/0110258}}].

\bibitem{Emparan:1999wa}
R.~Emparan, G.~T. Horowitz, and R.~C. Myers, {\it {`Exact description of black
  holes on branes'}},  JHEP {\bf 01} (2000) 007,
  [\href{http://arxiv.org/abs/hep-th/9911043}{{\tt hep-th/9911043}}].

\bibitem{Emparan:1999fd}
R.~Emparan, G.~T. Horowitz, and R.~C. Myers, {\it {`Exact description of black
  holes on branes 2. Comparison with BTZ black holes and black strings'}},
  JHEP {\bf 01} (2000) 021, [\href{http://arxiv.org/abs/hep-th/9912135}{{\tt
  hep-th/9912135}}].

\bibitem{Hubeny:2009ru}
V.~E. Hubeny, D.~Marolf, and M.~Rangamani, {\it {`Hawking radiation in large N
  strongly-coupled field theories'}},  Class.~Quant.~Grav. {\bf 27} (2010)
  095015, [\href{http://arxiv.org/abs/0908.2270}{{\tt arXiv:0908.2270}}].

\bibitem{Hubeny:2009kz}
V.~E. Hubeny, D.~Marolf, and M.~Rangamani, {\it {`Black funnels and droplets
  from the AdS C-metrics'}},  Class.~Quant.~Grav. {\bf 27} (2010) 025001,
  [\href{http://arxiv.org/abs/0909.0005}{{\tt arXiv:0909.0005}}].

\bibitem{Kinnersley:1970zw}
W.~Kinnersley and M.~Walker, {\it {`Uniformly accelerating charged mass in
  general relativity'}},  Phys.~Rev.~D {\bf 2} (1970) 1359.

\bibitem{Farhoosh:1981kc}
H.~Farhoosh and R.~L. Zimmerman, {\it {`Interior C-metric'}},  Phys.~Rev.~D
  {\bf 23} (1981) 299.

\bibitem{Bonnor:1983}
W.~B. Bonnor, {\it {`The sources of the C-metric'}},  Gen.~Rel.~Grav. {\bf 15}
  (1983) 535.

\bibitem{Dias:2002mi}
O.~J.~C. Dias and J.~P.~S. Lemos, {\it {`Pair of accelerated black holes in
  anti-de Sitter background: AdS C metric'}},  Phys.~Rev.~D {\bf 67} (2003)
  064001, [\href{http://arxiv.org/abs/hep-th/0210065}{{\tt hep-th/0210065}}].

\bibitem{Krtous:2003tc}
P.~Krtou\v{s} and J.~Podolsk\'{y}, {\it {`Radiation from accelerated black
  holes in de Sitter universe'}},  Phys.~Rev.~D {\bf 68} (2003) 024005,
  [\href{http://arxiv.org/abs/gr-qc/0301110}{{\tt gr-qc/0301110}}].

\bibitem{Krtous:2005ej}
P.~Krtou\v{s}, {\it {`Accelerated black holes in an anti-de Sitter universe'}},
   Phys.~Rev.~D {\bf 72} (2005) 124019,
  [\href{http://arxiv.org/abs/gr-qc/0510101}{{\tt gr-qc/0510101}}].

\bibitem{Hong:2003gx}
K.~Hong and E.~Teo, {\it {`A new form of the C metric'}},  Class.~Quant.~Grav.
  {\bf 20} (2003) 3269, [\href{http://arxiv.org/abs/gr-qc/0305089}{{\tt
  gr-qc/0305089}}].

\bibitem{Chen:2015vma}
Y.~Chen, Y.-K. Lim, and E.~Teo, {\it {`New form of the C metric with
  cosmological constant'}},  Phys.~Rev.~D {\bf 91} (2015) 064014,
  [\href{http://arxiv.org/abs/1501.01355}{{\tt arXiv:1501.01355}}].

\bibitem{Weyl1}
H.~Weyl, {\it {`Eine neue erweiterung der relativitätstheorie'}},  Ann.~Phys.
  {\bf 364} (1919) 101.

\bibitem{Weyl2}
H.~Weyl, {\it {`Reine Infinitesimalgeometrie'}},  Mathematische Zeitschrift
  {\bf 2} (1918) 384.

\bibitem{Bach1921}
R.~Bach, {\it {`Zur Weylschen relativit{\"a}tstheorie und der Weylschen
  erweiterung des Kr{\"u}mmungstensorbegriffs'}},  Mathematische Zeitschrift
  {\bf 9} (1921) 110.

\bibitem{Mannheim:1988dj}
P.~D. Mannheim and D.~Kazanas, {\it {`Exact Vacuum Solution to Conformal Weyl
  Gravity and Galactic Rotation Curves'}},  Astrophys.~J. {\bf 342} (1989) 635.

\bibitem{Mannheim:1992tr}
P.~D. Mannheim and D.~Kazanas, {\it {`Newtonian limit of conformal gravity and
  the lack of necessity of the second order Poisson equation'}},
  Gen.~Rel.~Grav. {\bf 26} (1994) 337.

\bibitem{Barabash:1999bj}
O.~V. Barabash and {\relax Yu}.~V. Shtanov, {\it {`Newtonian limit of conformal
  gravity'}},  Phys.~Rev.~D {\bf 60} (1999) 064008,
  [\href{http://arxiv.org/abs/astro-ph/9904144}{{\tt astro-ph/9904144}}].

\bibitem{Riegert:1984zz}
R.~J. Riegert, {\it {`Birkhoff's Theorem in Conformal Gravity'}},
  Phys.~Rev.~Lett. {\bf 53} (1984) 315.

\bibitem{1991PhRvD..44..417M}
P.~D. Mannheim and D.~Kazanas, {\it {`Solutions to the Reissner-Nordstr\"{o}m,
  Kerr, and Kerr-Newman problems in fourth-order conformal Weyl gravity'}},
  Phys.~Rev.~D {\bf 44} (1991) 417.

\bibitem{Brihaye:2009xc}
Y.~Brihaye and Y.~Verbin, {\it {`Cylindrically-Symmetric Solutions in Conformal
  Gravity'}},  Phys.~Rev.~D {\bf 81} (2010) 124022,
  [\href{http://arxiv.org/abs/0912.4669}{{\tt arXiv:0912.4669}}].

\bibitem{Verbin:2010tq}
Y.~Verbin and Y.~Brihaye, {\it {`Exact String-Like Solutions in Conformal
  Gravity'}},  Gen.~Rel.~Grav. {\bf 43} (2011) 2847,
  [\href{http://arxiv.org/abs/1008.1170}{{\tt arXiv:1008.1170}}].

\bibitem{Said:2012pm}
J.~L. Said, J.~Sultana, and K.~Z. Adami, {\it {`Charged Cylindrical Black Holes
  in Conformal Gravity'}},  Phys.~Rev.~D {\bf 86} (2012) 104009,
  [\href{http://arxiv.org/abs/1207.2108}{{\tt arXiv:1207.2108}}].

\bibitem{Liu:2012xn}
H.-S. Liu and H.~Lu, {\it {`Charged Rotating AdS Black Hole and Its
  Thermodynamics in Conformal Gravity'}},  JHEP {\bf 02} (2013) 139,
  [\href{http://arxiv.org/abs/1212.6264}{{\tt arXiv:1212.6264}}].

\bibitem{Klemm:1998kf}
D.~Klemm, {\it {`Topological black holes in Weyl conformal gravity'}},
  Class.~Quant.~Grav. {\bf 15} (1998) 3195,
  [\href{http://arxiv.org/abs/gr-qc/9808051}{{\tt gr-qc/9808051}}].

\bibitem{Cognola:2011nj}
G.~Cognola, O.~Gorbunova, L.~Sebastiani, and S.~Zerbini, {\it {`On the Energy
  Issue for a Class of Modified Higher Order Gravity Black Hole Solutions'}},
  Phys.~Rev.~D {\bf 84} (2011) 023515,
  [\href{http://arxiv.org/abs/1104.2814}{{\tt arXiv:1104.2814}}].

\bibitem{Lu:2012xu}
H.~Lu, Y.~Pang, C.~N. Pope, and J.~F. Vazquez-Poritz, {\it {AdS and Lifshitz
  Black Holes in Conformal and Einstein-Weyl Gravities}},  Phys. Rev. {\bf D86}
  (2012) 044011, [\href{http://arxiv.org/abs/1204.1062}{{\tt
  arXiv:1204.1062}}].

\bibitem{Lu:2013hx}
H.~Lü, Y.~Pang, and C.~N. Pope, {\it {Black Holes in Six-dimensional Conformal
  Gravity}},  Phys. Rev. {\bf D87} (2013), no.~10 104013,
  [\href{http://arxiv.org/abs/1301.7083}{{\tt arXiv:1301.7083}}].

\bibitem{Mannheim:2010ti}
P.~D. Mannheim and J.~G. O'Brien, {\it {`Impact of a global quadratic potential
  on galactic rotation curves'}},  Phys.~Rev.~Lett. {\bf 106} (2011) 121101,
  [\href{http://arxiv.org/abs/1007.0970}{{\tt arXiv:1007.0970}}].

\bibitem{Deliduman:2015vnu}
C.~Deliduman, O.~Kasikci, and B.~Yapiskan, {\it {`Flat Galactic Rotation Curves
  from Geometry in Weyl Gravity'}},
  \href{http://arxiv.org/abs/1511.07731}{{\tt arXiv:1511.07731}}.

\bibitem{Sultana:2012qp}
J.~Sultana, D.~Kazanas, and J.~L. Said, {\it {Conformal Weyl gravity and
  perihelion precession}},  Phys. Rev. {\bf D86} (2012) 084008.

\bibitem{Edery:1997hu}
A.~Edery and M.~B. Paranjape, {\it {`Classical tests for Weyl gravity:
  Deflection of light and radar echo delay'}},  Phys.~Rev.~D {\bf 58} (1998)
  024011, [\href{http://arxiv.org/abs/astro-ph/9708233}{{\tt
  astro-ph/9708233}}].

\bibitem{Edery:2001at}
A.~Edery, A.~A. Methot, and M.~B. Paranjape, {\it {`Gauge choice and geodetic
  deflection in conformal gravity'}},  Gen.~Rel.~Grav. {\bf 33} (2001)
  2075--2079, [\href{http://arxiv.org/abs/astro-ph/0006173}{{\tt
  astro-ph/0006173}}].

\bibitem{Sultana:2010zz}
J.~Sultana and D.~Kazanas, {\it {`Bending of light in conformal Weyl
  gravity'}},  Phys.~Rev.~D {\bf 81} (2010) 127502.

\bibitem{Cattani:2013dla}
C.~Cattani, M.~Scalia, E.~Laserra, I.~Bochicchio, and K.~K. Nandi, {\it
  {`Correct light deflection in Weyl conformal gravity'}},  Phys.~Rev.~D {\bf
  87} (2013) 047503, [\href{http://arxiv.org/abs/1303.7438}{{\tt
  arXiv:1303.7438}}].

\bibitem{Villanueva:2013gga}
J.~R. Villanueva and M.~Olivares, {\it {`On the Null Trajectories in Conformal
  Weyl Gravity'}},  JCAP {\bf 1306} (2013) 040,
  [\href{http://arxiv.org/abs/1305.3922}{{\tt arXiv:1305.3922}}].

\bibitem{Meng:2016gyt}
K.~Meng and L.~Zhao, {\it {`C-metric solution for conformal gravity with a
  conformally coupled scalar field'}},
  \href{http://arxiv.org/abs/1601.07634}{{\tt arXiv:1601.07634}}.

\bibitem{Liu:2010sz}
H.~Liu, H.~Lu, M.~Luo, and K.-N. Shao, {\it {`Thermodynamical Metrics and Black
  Hole Phase Transitions'}},  JHEP {\bf 12} (2010) 054,
  [\href{http://arxiv.org/abs/1008.4482}{{\tt arXiv:1008.4482}}].

\bibitem{Chen:2015zoa}
Y.~Chen, Y.-K. Lim, and E.~Teo, {\it {`Deformed hyperbolic black holes'}},
  Phys.~Rev.~D {\bf 92} (2015) 044058,
  [\href{http://arxiv.org/abs/1507.02416}{{\tt arXiv:1507.02416}}].

\bibitem{Harmark:2009dh}
T.~Harmark, {\it {`Domain Structure of Black Hole Space-Times'}},  Phys.~Rev.~D
  {\bf 80} (2009) 024019, [\href{http://arxiv.org/abs/0904.4246}{{\tt
  arXiv:0904.4246}}].

\bibitem{Mann:1996gj}
R.~B. Mann, {\it {`Pair production of topological anti-de Sitter black
  holes'}},  Class.~Quant.~Grav. {\bf 14} (1997) L109,
  [\href{http://arxiv.org/abs/gr-qc/9607071}{{\tt gr-qc/9607071}}].

\bibitem{Chen:2010zu}
Y.~Chen and E.~Teo, {\it {`Rod-structure classification of gravitational
  instantons with $U(1)\times U(1)$ isometry'}},  Nucl.~Phys.~B {\bf 838}
  (2010) 207, [\href{http://arxiv.org/abs/1004.2750}{{\tt arXiv:1004.2750}}].

\bibitem{Griffiths:2006tk}
J.~B. Griffiths, P.~Krtous, and J.~Podolsky, {\it {`Interpreting the
  C-metric'}},  Class.~Quant.~Grav. {\bf 23} (2006) 6745,
  [\href{http://arxiv.org/abs/gr-qc/0609056}{{\tt gr-qc/0609056}}].

\bibitem{Griffiths:2009dfa}
J.~Griffiths and J.~Podolsk{\'y}, {\em {`\textit{Exact space-times in
  Einstein's general relativity}'}}.
\newblock Cambridge University Press, (2009).

\end{thebibliography}\endgroup

\end{document}